\documentclass[prb, twocolumn, showpacs, superscriptaddress,longbibliography, floatfix]{revtex4-1}
\usepackage{amsmath, amsfonts, amssymb, mathrsfs}
\usepackage{graphicx, subfigure, color}
\usepackage{bm}
\usepackage{braket}
\usepackage{epstopdf}
\usepackage[normalem]{ulem} 
\usepackage{color}
\usepackage[breaklinks]{hyperref}
\usepackage{lipsum}
\usepackage[colorinlistoftodos]{todonotes}
\usepackage{xcolor}
\usepackage{enumerate}
\definecolor{grey}{rgb}{.6,.6,.6}

\hypersetup{
   colorlinks,
   citecolor=blue,
   filecolor=blue,
   linkcolor=blue,
   urlcolor=blue
}

\graphicspath{Graphes}

\begin{document}

\title{Braiding Majorana zero modes using quantum dots}

\author{Corneliu Malciu}
\affiliation{Laboratoire Pierre Aigrain, \'Ecole Normale Sup\'erieure-PSL Research University, CNRS, Universit\'e Pierre et Marie Curie-Sorbonne Universit\'es, Universit\'e Paris Diderot-Sorbonne Paris Cit\'e, 24 rue Lhomond, 75231 Paris Cedex 05, France}
\author{Leonardo Mazza}
\affiliation{D\'epartement de physique, \'Ecole Normale Sup\'erieure-PSL Research University, CNRS, 24 rue Lhomond, 75231 Paris Cedex 05, France} 
\author{Christophe Mora}
\affiliation{Laboratoire Pierre Aigrain, \'Ecole Normale Sup\'erieure-PSL Research University, CNRS, Universit\'e Pierre et Marie Curie-Sorbonne Universit\'es, Universit\'e Paris Diderot-Sorbonne Paris Cit\'e, 24 rue Lhomond, 75231 Paris Cedex 05, France}

\begin{abstract}
We discuss a network of Kitaev wires coupled to several individually-tunable quantum dots as an extension of the recent experiments on a quantum dot coupled to a nanowire hosting Majorana zero modes [Deng et al.\ Science \textbf{354} 1557 (2016) and Deng et al.\ arXiv:1712.03536 (2017)]. 
The setup features localized Majorana modes with exact zero energy and we show that they can be manipulated by solely acting on the quantum dots. 
A braiding process can be obtained by arranging three wires as a trijunction and a charge readout of the quantum dots can be used to reveal the non-Abelian statistics of Majorana zero modes. 
The setup can be scaled up to serve the more advanced purposes of topological quantum computation.
\end{abstract}

\date{\today}
\maketitle

\section{Introduction}

After their introduction in 1937 in the context of the relativistic Dirac equation~\cite{Majorana_1937}, Majorana fermions have recently experienced a renewed interest for their relevance in the description of some low-dimensional and superconducting topological models~\cite{Alicea_2012, Leijnse_2012, Beenakker_2013, Stanescu_2013, Elliott_2015, DasSarma_2015,   Masatoshi_2016, Aguado_2017}. In this context, they typically bind to defects (e.g.\ vortices in the p+ip model~\cite{ReadGreen2000}) or to boundaries between topological and non-topological phases (e.g.\ at the edges of Kitaev's chain~\cite{Kitaev_2001}); they are zero-energy modes whose appearance is topologically protected from small perturbations. Remarkably, Majorana zero modes (MZM) exhibit an exchange statistics that is neither bosonic nor fermionic: they are non-Abelian anyons and as such they imply a degeneracy of the many-body ground state~\cite{Ivanov_2001}.

The topological protection of MZMs gave them a very special status: they lie at the heart of current proposals for hardware-based fault-tolerant quantum computation~\cite{Nayak_2008, Bonderson_Freedman_2008, Bonderson_Freedman_2009, Sau_Tewari_Sumanta_2010, Pachos_book_2012, Mazza_Rizzi_2013, Hyart_Beenakker_2013,  Bonderson_2013, Aasen_Hell_2016,  Clarke_Sau_Jay_2016,  Ippoliti_Rizzi_2016}. Their non-Abelian statistics can be used to perform non-trivial operations on the ground states through the adiabatic exchange of two anyon positions, which is described by their braiding group~\cite{Alicea_Oreg_2011, Clarke_Sau_2011, Sau_Tewari_2011, Cheng_Galitski_2011, Romito_Alicea_2012, Halperin_Oreg_2012, Heck_Beenakker_2012, Burrello_Heck_2013, Scheurer_Shnirman_2013, Ching-Kai_Vazifeh_2015, Karzig_Pientka_2015, Amorim_Ebihara_2015, Bauer_Karzig_2018}. 
Since it is not sufficient for performing universal quantum computation, it has been suggested that projective measurements could be used to implement the missing gates\cite{Bravyi_kitaev_2002, Bravyi_kitaev_2005, Bravyi_2006}. 

There have been important experimental progresses in the recent years regarding the realization and observation of MZMs in solid-state devices. The community has focused on hybrid semiconductor-superconductor setups~\cite{Fu_Kane_2008, Oreg_Refael_2010, Lutchyn_Sau_2010, Mourik_2012, Das_2012, Fulga_Haim_2013, Churchill_2013, Finck_2013, NadjPerge_2014, Albrecht_2016, Deng_2016, Deng_2017, Lutchyn_2017} and chains of magnetic adatoms coupled to conventional superconductors~\cite{choy2011,NadjPerge2013,klinovaja2013,yazdani2014,ruby2015,pawlak2016,feldman2017}. The recent proposals for MZMs in two-dimensional electron gases proximitized to superconductors promise a new generation of platforms significantly less affected by disorder~\cite{Hell_Leijnse_2017, Hell_Flensberg_2017}. Yet, discerning MZMs from the variety of phenomena producing sub-gap states is still an experimental challenge~\cite{Sau_Tewari_DasSarma_2011, DasSarma_Nag_2016}.

In order to distinguish topological and trivial excitations without performing braiding, it has been suggested that a quantum dot (QD) could be used as a probe of the non-locality of MZM~\cite{Clarke_2017, Prada_San-Jose_2017, Liu_Sau_2017, Ptok_2017, Schuray_Weithofer_2017, Liu_Sau_2018, Chevallier_Szumniak_2018}. The key idea is to alter the zero-bias conductance measurements by adding a tunable QD between the lead and the wire: 
the entire conductance pattern observed as the dot is tuned through resonance provides information about topological properties. The successful experimental realization of such a device and the characterization of the non-local nature of the zero-energy excitation~\cite{Deng_2016, Deng_2017} demonstrate that nanowires coupled to QDs are an experimentally practical and promising route for the study and exploit of electronic liquids supporting Majorana zero modes.

In this work, we show that such experimental setups can also be employed for performing braiding operations. 
We demonstrate this by studying a network of Kitaev wires, each one connected to one or more QDs, and show that they host exact MZMs that can be controlled and transported at will by only manipulating the QDs.
Additionally, the QDs serve the purpose of parity readout~\cite{Gharavi_Baugh_2016}, which is necessary for obtaining the experimental confirmation that the braiding process occurred.
By measuring the parity of the total charge of two neighbouring QDs, and considering larger networks with a more complex structure, it is possible to scale up the system with the objective of topological quantum computation.

This article is organized as follows: in Sec.~\ref{sect:QD-wire} we briefly recall some results on Kitaev's wire coupled to a QD.
In Sec.~\ref{sect:braiding_mill} we discuss how to braid the MZMs that appear in such setup and how to measure the outcome of such operation using QDs.
In Sec.~\ref{sec:tqc} we outline the scale-up of such protocol to the realization of topological quantum computation in QD-controlled circuits.
Our conclusions are presented in Sec.~\ref{sect:conclusion}.

\section{A Kitaev wire coupled to a quantum dot}\label{sect:QD-wire}

In this section we consider a tunable QD which is tunnel-coupled to the left edge of a Kitaev wire, a setup that has been theoretically studied in Ref.~\onlinecite{Clarke_2017}, and also in the more experimentally-relevant spinful case in Ref.~\onlinecite{Prada_San-Jose_2017}.
We briefly review some of their results and stress that thanks to the tunability of the QD  it is always possible to have Majorana modes with exactly zero energy.

\subsection{Effective model}

\begin{figure}
    \centering
    \includegraphics[width =\columnwidth]{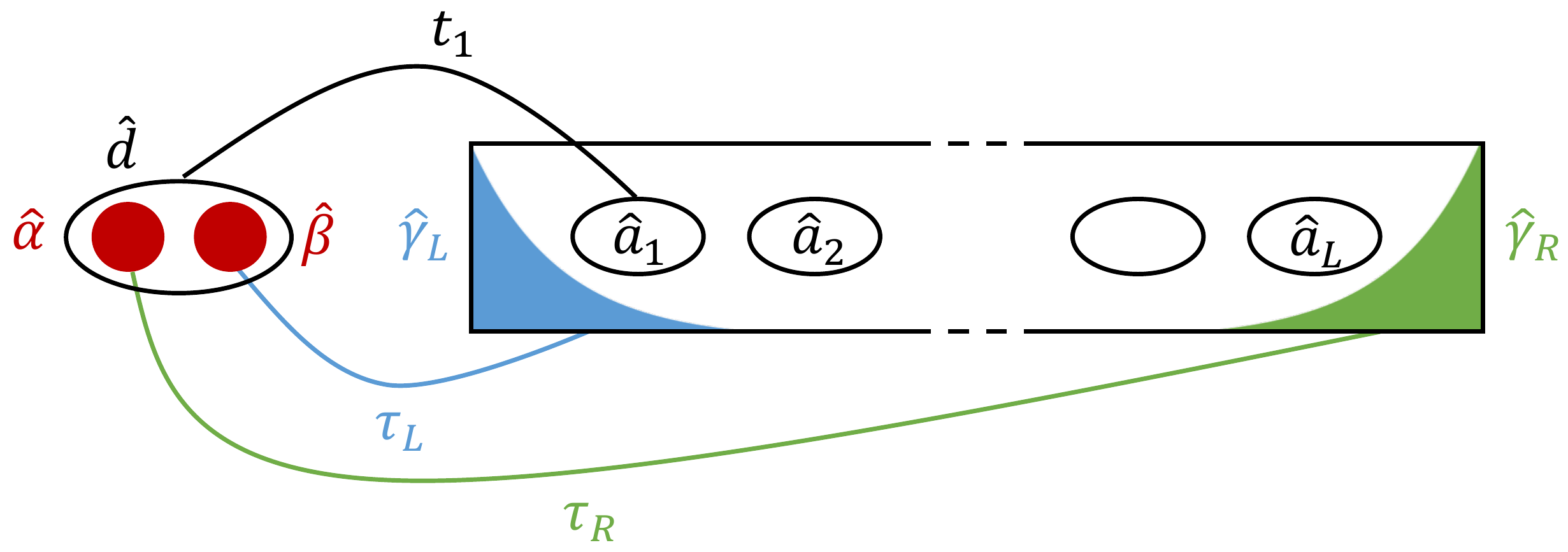}
    \caption{Scheme of the QD-wire setup: a QD with a single fermionic mode is connected to the first site of a Kitaev wire through a real hopping amplitude~$t_1$. The low-energy subspace is described by four Majorana modes $\hat\alpha$, $\hat\beta$ (QD) and $\hat\gamma_L$, $\hat\gamma_R$ (wire, exponentially localized at the edges). The effective couplings $\tau_{L,R}$ between the QD and the wire are defined in Eq.~\eqref{eq:H_QD-K_eff} and shown in the figure.}
    \label{fig:scheme_model}
\end{figure}

We model the QD with a single fermionic mode using canonical operators $\hat d^{(\dagger)}$; the Kitaev wire has length $L$ and its fermionic modes are described by the operators $\hat a_{j}^{(\dagger)}$, $j = 1 \ldots L$.
The full Hamiltonian reads (see Fig.~\ref{fig:scheme_model}):
\begin{align}\label{eq:H_QD-K_micro}
    &\hat H_{\text{QD-K}} = e_d \, \hat{d}^\dagger \hat{d} - t_1 \left( \hat{d}^\dagger \hat{a}_1 + \hat{a}_1^\dagger \hat{d} \right) + \hat H_K; \\
    \nonumber
    &\hat H_K = \sum_{j=1}^L \left\{ \left( -t \hat{a}^\dagger_j \hat{a}_{j+1} - \Delta \hat{a}_j \hat{a}_{j+1} + \text{h.c.} \right) - \mu \, \hat{a}^\dagger_j \hat{a}_{j} \right\};
\end{align}
where $e_d$ is the tunable energy of the QD, $t$ is the hopping term of the wire, $\Delta = |\Delta|e^{i\phi}$ its pairing term and $\mu$ its chemical potential. Without loss of generality $t_1$ is taken real.
For $|\mu|<2t$ and $\Delta \neq 0$, Hamiltonian $\hat H_K$ exhibits topological subgap states, which are Majorana modes exponentially localized at the left and right edges of the wire, $\hat{\gamma}_L$ and $ \hat{\gamma}_R$; for a finite length $L$, they have an exponentially-small energy  $\varepsilon\sim e^{-L}$. We assume to be in this topological phase.

In order to capture the low-energy behaviour of the full system, it is sufficient to consider the coupling of the QD to the edge states of the wire. We perform a gauge transformation on the $\hat d^{(\dagger)}$ so that,
by decomposing it into two Majorana fermions $\hat{\alpha} = \hat d + \hat d^\dagger, \hat{\beta} = -i (\hat d - \hat d^\dagger)$, we get (see the sketch in Fig.~\ref{fig:scheme_model}):
\begin{equation}\label{eq:H_QD-K_eff}
    \hat H_{\text{QD-K}}^{\text{eff}} = i \left( \xi \,\hat{\alpha} \hat{\beta} + \varepsilon \, \hat{\gamma}_L\hat{\gamma}_R + \tau_L \, \hat{\beta} \hat{\gamma}_L + \tau_R \hat{\alpha}\hat{\gamma}_R \right),
\end{equation}
where $\xi$ = $e_d/2$. Terms proportional to the identity have been omitted and the expression is particularly simple because Hamiltonian~\eqref{eq:H_QD-K_micro} is unitarily related to a time-reversal invariant one.
The effective couplings $\tau_L$ and $ \tau_R $ depend on the projection of $\hat{\gamma}_L $ and $ \hat{\gamma}_R$ on the first site $\hat a_1$ of the wire, and can be chosen real; clearly, $\tau_R\sim e^{-L}$. Because this Hamiltonian describes the sub-gap physics, we implicitly assumed $\xi, \tau_L, \tau_R, \varepsilon \ll E_G$, where $E_G \sim |\Delta|$ is the energy gap of the many-body system.

\begin{figure}
    \centering
    \includegraphics[width = \columnwidth]{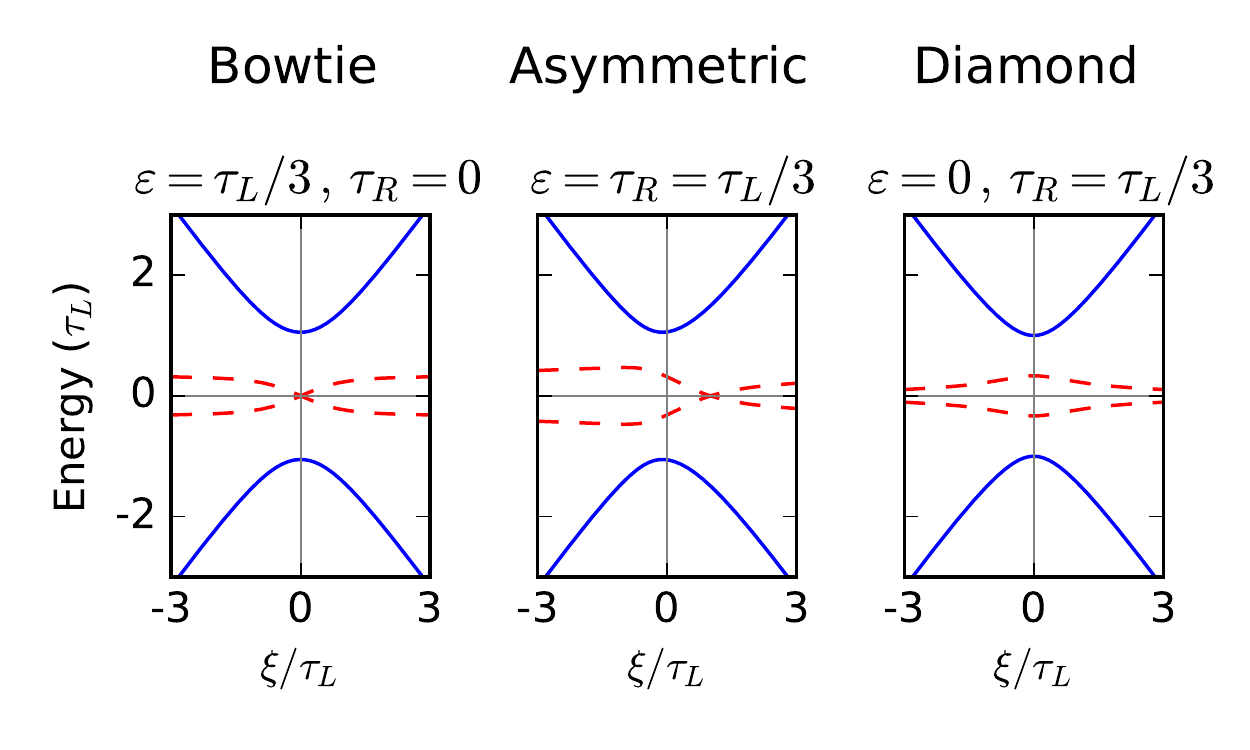}
    \caption{The \textit{bowtie}, \textit{asymmetric} and \textit{diamond} patterns observed in the model~\eqref{eq:H_QD-K_eff} when the QD energy $\xi$ crosses zero. Far from $\xi = 0$, the linear dispersion (solid blue) corresponds to the QD energies, and the split levels (dashed red) correspond to the $\hat \gamma_{L,R}$ of the wire. The patterns are observed both with the effective model (plotted here) and with the full microscopic Hamiltonian~\eqref{eq:H_QD-K_micro} (not shown). Assuming $|\tau_R| \ll |\tau_L|$, couplings are extracted as follows: $\varepsilon$ is the energy splitting of the Majorana $\hat \gamma_{L,R}$ far from resonance, $2 |\tau_L|$ is the minimum gap between the solid blue lines, the position of this minimum sets the origin for $\xi$, and finally $\tau_R$ is  obtained from the zero-energy crossing predicted by Eq.~\eqref{eq:resonance}. }
    \label{fig:eff_model_patterns}
\end{figure}

Three different patterns, dubbed \textit{bowtie} $(\varepsilon\ne 0,\tau_R = 0)$, \textit{asymmetric} $(\varepsilon\ne 0,\tau_R \ne 0)$, and \textit{diamond} $(\varepsilon=0, \tau_R \ne 0)$ can be observed and are shown in Fig.~\ref{fig:eff_model_patterns}. They exhibit a zero-energy crossing, indicating a parity switch of the many-body ground state, occurring exactly at (see Appendix~\ref{app:parity_switch}):
\begin{equation}
    \xi_c = -\frac{\tau_L\tau_R}{\varepsilon}.
    \label{eq:resonance}
\end{equation}
For the fine-tuned \textit{diamond} case where the energy splitting $\varepsilon$ between the two Majorana bound states vanishes, $|\xi_c| \to \infty$.
All three patterns can be recovered from the microscopic Hamiltonian in Eq.~\eqref{eq:H_QD-K_micro}. The effective couplings used in Eq.~\eqref{eq:H_QD-K_eff} can be extracted through spectroscopic (non-linear conductance) measurements as recently observed by Deng et al.~\cite{Deng_2016, Deng_2017}.

The parity switch of the ground state occurring at $\xi_c$ can be understood in terms of charge transport from the dot to the wire. As $\xi$ is adiabatically tuned from $-\infty$ to $+\infty$ during a period $T$ (sent to $+\infty$ for a truly adiabatic evolution), the QD which was initially charged unloads into the wire. At $t = 0$ and $t =T$ the dot and the wire are effectively uncoupled, and both subsystems have a definite parity:
\begin{equation}
    \begin{array}{ccc}
        t=0\: : \quad & \hat P_d = i\hat{\alpha} \hat{\beta} = 1 \quad & \hat P_w = i\hat{\gamma}_L \hat{\gamma}_R = \chi \\ \\
        t=T\: : \quad & \hat P_d = - 1 \quad & \hat P_w = -\chi 
    \end{array}
\end{equation}
where  $\chi = - {\rm sgn} (\varepsilon)$. Because the total parity $\hat P = \hat P_d \hat P_w$ is conserved by the Hamiltonian, the final wire parity must be $-\chi$. 
This protocol adiabatically drives the system into an excited state similarly to what occurs in the $4\pi$ Josephson effect. Note, however, that $\hat P_w$ is reversed here, whereas it remains unchanged in the Josephson setup.

\subsection{Exact MZMs}\label{sect:exact_MZMs}

In the most general case, a finite Kitaev wire does not host true zero-energy states (called MZMs) since $\varepsilon \neq 0$. 
However,  the effective model~\eqref{eq:H_QD-K_eff} is degenerate for $\xi = \xi_c$, and therefore hosts a pair of MZMs, denoted below  $\hat \gamma_1$ and $\hat \gamma_2$ (MZMs always come in pairs because of the particle-hole symmetry of the Bogoliubov-de Gennes Hamiltonian).
Thus, a tunnel contact to a quantum dot with a tunable orbital energy can induce genuine MZM, later needed for topologically-protected quantum operations.

In order to gain further insights on the spatial position of $\hat \gamma_{1}$ and $\hat \gamma_2$, we first consider the simplest case $\varepsilon = 0$ and $\tau_R = 0$, so that the Hamiltonian
\begin{equation}
    \hat H_{\text{QD-K}}^{\text{eff}} = i  \xi \,\hat{\alpha} \hat{\beta} + i \tau_L \, \hat{\beta} \hat{\gamma}_L,
\end{equation}
decouples the Majorana fermion $\hat \gamma_R$: 
$\hat \gamma_R $ is a MZM and we arbitrarily identify it with $ \hat \gamma_2$.  
The other MZM is:
\begin{equation} \label{eq:mzm_qd-wire_topo}
  \hat \gamma_1 = \frac{1}{\mathcal{N}}\left( \hat \gamma_L + \frac{\tau_L}{\xi} \hat \alpha \right), \qquad \mathcal{N} = \sqrt{1+\left( \frac{\tau_L}{\xi} \right) ^2},
\end{equation}
whose spatial localization is controlled by the QD energy $\xi$; for instance  $\hat\gamma_1 = \hat \gamma_L$ when $|\xi|\rightarrow\infty$ and  $\hat \gamma_1 = \hat \alpha$ when $\xi = 0$.
This ability to move the MZM from the wire to the QD by tuning the QD energy $\xi$ is the key to the braiding and readout procedures presented in this article. 
Note also the spatial separation of $\hat \gamma_1$ and $\hat \gamma_2$.

In the general case, namely the \textit{asymetric} configuration in Fig.~\ref{fig:eff_model_patterns}, the effective Hamiltonian~\eqref{eq:H_QD-K_eff} exhibits two MZM at $\xi = \xi_c$ given by
 \begin{subequations}
\label{eq:mzm_qd-wire_bowtie}
\begin{align}
  \label{eq:mzm_qd-wire_bowtie_1} \hat{\gamma_1} &= \frac{1}{\mathcal{N}_{1}} \left( \hat \alpha - \frac{\tau_R}{\varepsilon}\hat{\gamma}_L\right), \qquad  \mathcal{N}_{1} = \sqrt{1+\left( \frac{\tau_R }{\varepsilon}\right) ^2},\\
    \label{eq:mzm_qd-wire_bowtie_2}
    \hat{\gamma}_2 &= \frac{1}{\mathcal{N}_2} \left(\hat{\gamma}_R + \frac{\varepsilon}{\tau_L}\hat{\beta} \right), \qquad \mathcal{N}_2 = \sqrt{1+\left( \frac{\varepsilon }{\tau_L }\right) ^2}.
\end{align}
\end{subequations}
In the \textit{bowtie} configuration, $\tau_R$ vanishes and thus $\xi_c=0$. Setting the QD energy to zero $\xi=0$ in the Hamiltonian~\eqref{eq:H_QD-K_eff} decouples $\hat \alpha$ and thus localizes the first MZM on the QD ($\hat \gamma_1 = \hat \alpha$); this is in agreement with Eq.~\eqref{eq:mzm_qd-wire_bowtie_1}.

For simplicity, we will set $\tau_R=0$ in the remainder of this paper. For non-zero $\tau_R$, the localization of the first MZM is not exactly on the QD (still it remains in the vicinity), as given by Eq.~\eqref{eq:mzm_qd-wire_bowtie_1}, but $\xi = \xi_c$ still enforces MZMs.

\section{Braiding MZMs with QDs}\label{sect:braiding_mill}

In this section we discuss how to braid the MZMs that we identified above. We first present a generic argument concerning the adiabatic transport of MZMs and then explicitly apply it to a trijunction where their non-Abelian statistics can be revealed through braiding.

\subsection{Transporting Majorana fermions}\label{sect:transport}

We employ an argument due to Kitaev (see Sau \textit{et al.}\cite{Sau_Tewari_2011}) in order to show how topologically-protected transport of Majorana fermions can be achieved with tunable QDs. By topological protection, we mean that the final state after an adiabatic evolution is only dictated by the final and initial conditions and does not depend on the intermediate details. 

\begin{figure}
    \centering
    \includegraphics[width = \columnwidth]{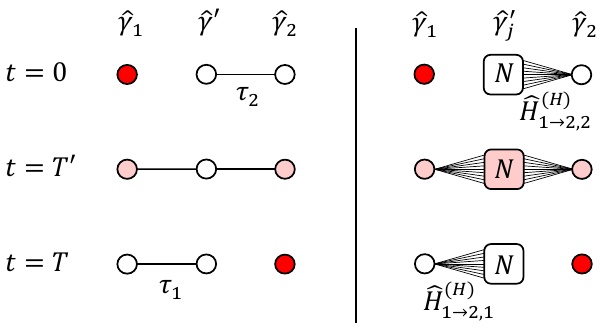}
    \caption{
    \textit{Left:} Transport of a MZM from $\hat \gamma_1$ to $\hat \gamma_2$ through $\hat \gamma'$ by tuning the couplings $\tau_{1,2}$. Because the total number of Majorana fermions is odd, a MZM must exist at all times. At $t=0$, $\tau_1 = 0$ and the MZM is strictly localized on $\hat \gamma_1$; similarly, at $t=T$, $\tau_2 = 0$ and the MZM is strictly localized on $\hat \gamma_2$; at intermediary times $t=T'$, the MZM is delocalized among $\hat \gamma_1$ and $\hat \gamma_2$. Provided that the couplings are tuned adiabatically, and that the gap of the higher energy levels does not close, $\hat\gamma_2(T) = \pm\hat \gamma_1(0)$ which corresponds to the transport $1\rightarrow2$. 
    \textit{Right:} Generalization of the protocol amounting to the transport of a MZM from $\hat \gamma_1$ to $\hat \gamma_2$ through a network of $N$ Majorana fermions $\hat \gamma'_j$ ($N$ odd). Once again, the total number of Majorana fermions is odd, ensuring the existence of a MZM at all times.
    }
    \label{fig:transport}
\end{figure}

We begin with a simple example by considering three Majorana fermions $\hat \gamma_{1}$, $\hat \gamma_2$ and $\hat \gamma'$ connected by time-dependent couplings:
\begin{equation}\label{eq:H_transport}
    \hat H_{1\rightarrow2}(t) = i \tau_1(t) \hat \gamma_1 \hat \gamma' + i \tau_2(t) \hat \gamma_2 \hat \gamma'
\end{equation}
and represented in Fig.~\ref{fig:transport}, left panel. A fourth Majorana mode is implicitly assumed to be at zero energy, uncoupled and unperturbed during the whole time evolution. As discussed in Sec.~\ref{sect:exact_MZMs}, particle-hole symmetry dictates that MZMs come in pair, and therefore there is an additional MZM which is a combination of $\hat \gamma_1$, $\hat \gamma_2$ and $\hat \gamma'$ and the ground state is twofold degenerate  at all times.
The procedure starts at time $t=0$ with $\tau_1(0)=0$ and the MZM localized on $\hat \gamma_1$; it ends at time $t=T$ with $\tau_2(T)=0$ and the MZM localized on $\hat \gamma_2$ (see Fig.~\ref{fig:transport}, left panel). We assume that the two higher energy states are gapped out during the entire process, and that the couplings $\tau_{1,2}(t)$ are tuned adiabatically, so that the system remains in the ground state of Hamiltonian~\eqref{eq:H_transport}.

We discuss time evolution in the Heisenberg picture and use the superscript $(H)$ for Heisenberg operators, including the Hamiltonian. 
Since the Hamiltonian $\hat H^{(H)}_{1 \to 2}(t)$ and the parity operator\footnote{In fact,   $\hat P^{(H)}(t) = i \hat \gamma_{1}(t)\hat \gamma_2(t)\hat \gamma'(t)$ is not the fermionic parity. Nonetheless the same proof holds if one replaces it with the true parity operator $\hat P^{(H)}_2(t) = - \hat \gamma_0(t) \hat \gamma_{1}(t)\hat \gamma_2(t)\hat \gamma'(t)$ as $\hat \gamma_0(t)$ commutes with the Hamiltonian at all times} $\hat P^{(H)}(t) = i \hat \gamma_{1}(t)\hat \gamma_2(t)\hat \gamma'(t)$ commute at all times, it holds that $\hat P^{(H)}(T)= \hat P^{(H)}(0)$. 
Adiabatic evolution implies that an initial ground state $\ket{\Psi_0}$ of $\hat H_{1\rightarrow2}(0)$ remains a ground state of  $\hat H^{(H)}_{1 \to 2}(t)$ at all times. Hence we have, at initial and final times,
\begin{equation}\label{rel2}
  \begin{split}
  i \hat \gamma_2 \hat \gamma'
 \ket{\Psi_0} & =  - \text{sgn}[\tau_2(0)] \ket{\Psi_0}, \\[2mm]
  i  \hat \gamma_1 (T) \hat \gamma' (T)
  \ket{\Psi_0} & = -\text{sgn}[\tau_1(T)]  \ket{\Psi_0}.
  \end{split}
\end{equation}
Combining this result with parity conservation, we find:
\begin{equation}\label{eq:transport}
    \hat{\gamma_2}(T) \ket{\Psi_0} =  \pm \hat{\gamma}_1(0) \ket{\Psi_0};
\end{equation}
where the sign depends on the microscopic details of the Hamiltonian (here, it is $\text{sgn}[-\tau_1(T)\tau_2(0)]$).
Eq.~\eqref{eq:transport} is interpreted as the adiabatic transport of a MZM from $\hat{\gamma}_1$ to $\hat{\gamma}_2$. Note that the same proof holds if the system is initialized in one of the two excited states $\ket{\Psi_e}$, and Eq.~\eqref{eq:transport} is therefore true at the operatorial level, i.e. $\hat{\gamma_2}(T) = \pm \hat{\gamma}_1(0)$.

This scheme can be generalized to the case of the transport of a MZM from $\hat \gamma_1$ to $\hat \gamma_2$ through an arbitrary network  involving an odd number of Majorana fermions $\hat \gamma'_j$, $j=1\hdots N$, where $N$ odd ensures the existence of a MZM (see Fig.~\ref{fig:transport}, right panel). Again, we are implicitly assuming the existence of an additional MZM $\hat\gamma_0$ which never appears in the Hamiltonian but ensures an even number of Majorana modes in the system. 
The Hamiltonian reads:
\begin{align}
    \hat H_{1\rightarrow2}(\{\hat \gamma_1, &\hat \gamma_2, \hat \gamma'_j\}, t) = \hat H_{1\rightarrow2,1}(\{\hat \gamma_1, \hat \gamma'_j\}, t) +\\
    \nonumber
    & + \hat H_{1\rightarrow2,2}(\{\hat \gamma_2, \hat \gamma'_j\}, t) 
    + \hat H_{1\rightarrow2,3}(\{\hat \gamma'_j\}, t).
\end{align}
The first (second) term describes the coupling of the MZM $\hat \gamma_1$ ($\hat \gamma_2$) to the Majorana modes of the network $\{\hat \gamma'_j\}$. 
The third term is the Hamiltonian of the network. Note that none of these operators is assumed to be bilinear in the Majorana modes (they could also be quartic); however, they all conserve the total fermionic parity.
Thus, $[\hat H_{1\rightarrow2}^{(H)}(t), \hat P^{(H)}(t)]=0$ at all times, where $\hat P^{(H)}(t) = \hat \gamma_1(t) \hat \gamma_2(t) \prod_j \hat \gamma'_j(t)$.

At $t=0$, $\hat H_{1\rightarrow2,1}=0$ and $\hat \gamma_1$ is uncoupled. The two MZMs $\hat\gamma_0$ and $\hat \gamma_1$ commute with the full Hamiltonian $ \hat H_{1\rightarrow2}$ and thus generate the twofold degenerate ground state. We define the partial parity operator $\hat P_{e,1} =\hat \gamma_2 \prod_j \hat \gamma'_j$ commuting with the Hamiltonian $\hat H_{1\rightarrow2}$, $\hat\gamma_0$ and $ \hat\gamma_1$.
It follows that any ground state $\ket{\Psi_0}$ of $\hat H_{1\rightarrow2}$ is an eigenstate of  $\hat P_{e,1}$ with eigenvalue  $\chi_1 = \pm 1$. The sign of $\chi_1$ depends specifically on the microscopic details of $\hat H_{1 \rightarrow2 }$ but remains the same within the ground state subspace.
In the particular case of quadratic Hamiltonians,  $\chi_1$ is obtained by computing the sign of the Pfaffian of the matrix defined from the Majorana pairwise couplings~\cite{Kitaev_2001}.
Summarizing:
\begin{equation}
    \hat P^{(H)} (0)\ket{\Psi_0} = \hat \gamma_1 \hat P_{e,1} \ket{\Psi_0} = \chi_1 \hat \gamma_1 \ket{\Psi_0}.
\end{equation}

At time $t=T$, $\hat \gamma_2(T)$ is uncoupled since $\hat H^{(H)}_{1\rightarrow2,2}(T)=0$.
Since during the entire process there are two MZMs, the ground state remains twofold degenerate. An adiabatic evolution means that the initial ground state $\ket{\Psi_0}$ is also a ground state of $\hat H_{1\rightarrow2}^{(H)}(T)$. Repeating the same arguments as for $t=0$, we introduce the parity operator  $\hat P_{e,2}^{(H)}(T)=\hat \gamma_1(T) \prod_j \hat \gamma'_j(T)$ commuting with the final Hamiltonian  $\hat H_{1\rightarrow2}^{(H)}(T)$ as well as with the MZMs $\hat \gamma_0(T) = \hat \gamma_0$ and $\hat \gamma_2(T)$. One obtains:
\begin{equation}
    \hat P^{(H)}(T)\ket{\Psi_0} = -\hat \gamma_2(T)\hat P_{e,2}^{(H)}(T)\ket{\Psi_0} = -\chi_2 \hat \gamma_2(T) \ket{\Psi_0},
\end{equation}
where $\chi_2$ is the eigenvalue of  $\hat P_{e,2}^{(H)}(T)$ over the ground state subspace.

The parity conservation reads $\hat P^{(H)}(T) = \hat P^{(H)}(0)$ leading to:
\begin{equation}\label{eq:braiding_sign}
    \hat \gamma_2(T) \ket{\Psi_0} = - \chi_1 \chi_2 \hat \gamma_1(0) \ket{\Psi_0},
\end{equation}
which corresponds to a MZM transport at the ground state level, and where $\chi_{1,2}$ can be explicitly computed for any specific example. If the process is adiabatic with respect to all possible energy differences in the spectrum, the same derivation holds for an arbitrary excited state $\ket{\Psi_e}$.

\begin{figure*}
    \centering
    \includegraphics[width = \textwidth]{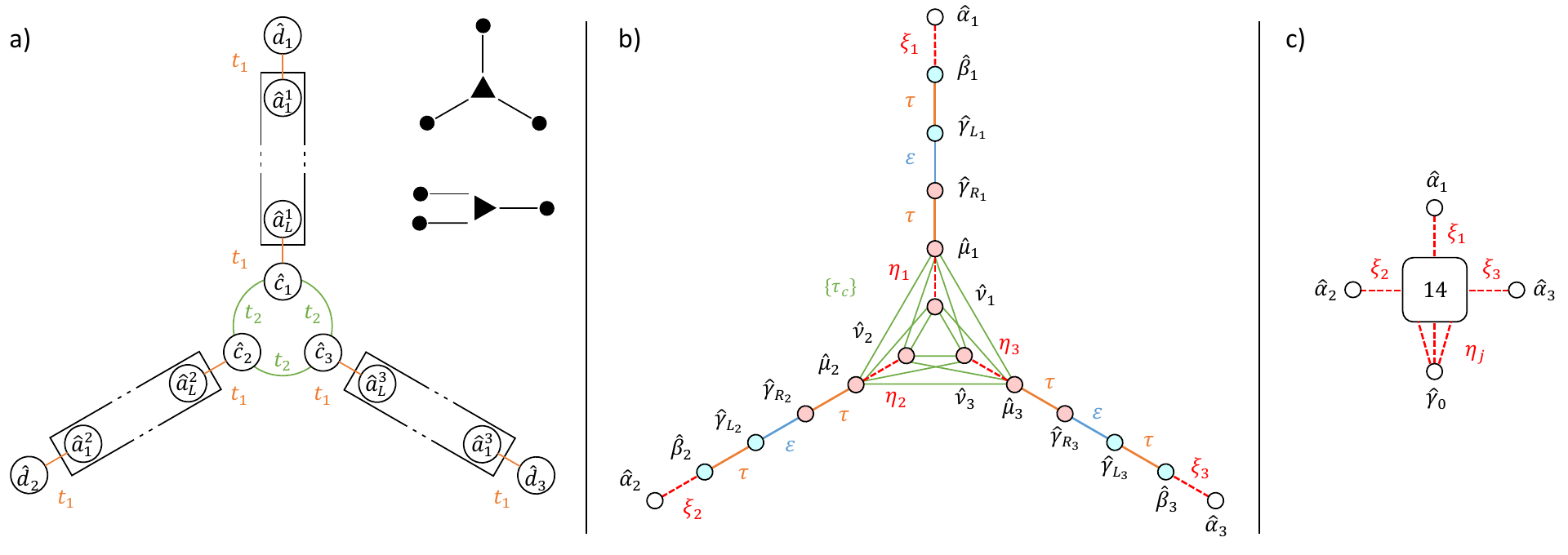}
    \caption{
    a) The Majorana trijunction. It is built by connecting three identical QD-wire-QD systems through a hopping term $t_2$. 
    A compact representation in given in the upper right corner, where a line stands for a Kitaev wire, a circle for a QD, and a triangle for 3 QDs connected by hopping terms; note that the trijunction we propose can in principle be realized with parallel nanowires. 
    b) The low-energy behaviour of this system is captured by a set of $18$ Majorana fermions (see Eq.~\eqref{eq:H_mill2_MF}). 
    c) The entire system amounts to $4$ MZMs $\hat \alpha_{1}$, $\hat\alpha_2$, $\hat\alpha_3$ and $ \hat \gamma_0$ connected to a network of $14$ Majorana fermions by tunable couplings $\xi_j, \eta_j$.
    }
    \label{fig:braiding_mill_inner_dots}
\end{figure*}

\subsection{Majorana trijunction}\label{sect:trijunction}

\begin{figure}
    \centering
    \includegraphics[width =\columnwidth]{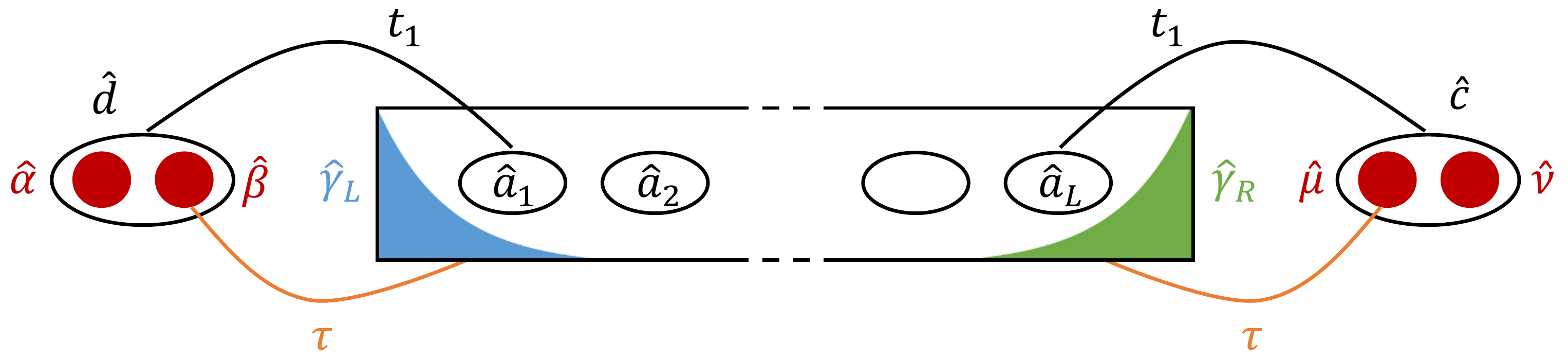}
    \caption{Scheme of a QD-wire-QD setup: two QDs with a single fermionic mode are connected to the edges of a Kitaev wire through a real hopping amplitude~$t_1$. The low-energy subspace is described by six Majorana modes $\hat\alpha$, $\hat\beta$ (left QD), $\hat\mu$, $\hat\nu$ (right QD) and $\hat\gamma_L$, $\hat\gamma_R$ (wire, exponentially localized at the edges). The effective couplings $\tau$ between the QD and the wire are defined in Eq.~\eqref{eq:H_QD-K-QD_eff} and shown in the figure.}
    \label{fig:scheme_qd_k_qd}
\end{figure}

We now apply the generic principles of adiabatic transport of MZMs outlined in the previous section to a trijunction~\cite{Sau_Tewari_2011, Alicea_Oreg_2011, Heck_Beenakker_2012, Halperin_Oreg_2012, Burrello_Heck_2013}, the simplest setup for braiding MZMs. The trijunction that we develop here braids MZMs by only tuning QD energy levels, and its elementary constituent is the QD-wire-QD setup displayed in Fig.~\ref{fig:scheme_qd_k_qd} and described by the Hamiltonian:
\begin{equation}\label{eq:H_QD-K-QD_micro}
	\hat H_{\text{QD-K-QD}} = \hat H_{\text{QD-K}} + e_d' \hat c^\dagger \hat c - t_1 \left(\hat{a}_L^\dagger \hat{c} + \hat{c}^\dagger \hat{a}_L\right).
\end{equation}
Here, $\hat H_{\text{QD-K}}$ is in Eq.~\eqref{eq:H_QD-K_micro}, $e_d'$ is the energy level of the right QD, and $\hat{c}^{(\dagger)}$ is the fermionic annihilation (creation) operator on this QD. For simplicity, both QDs are connected to the wire by the same hopping term $t_1$. The associated low-energy effective model in the \textit{bowtie} configuration is:
\begin{equation}\label{eq:H_QD-K-QD_eff}
\hat H_{\text{QD-K-QD}}^{\text{eff}} = i \left( \xi \hat{\alpha} \hat{\beta} + \eta \hat \mu \hat\nu + \varepsilon \hat{\gamma}_L\hat{\gamma}_R + \tau \, \hat{\beta} \hat{\gamma}_L + \tau \, \hat{\mu} \hat{\gamma}_R\right)
\end{equation}
where $\hat{\mu}$ and $\hat{\nu}$ are the Majorana fermions of the right QD, and where $\eta = e_d'/2$. Note that we have changed the notation $\tau_L$ used in Eq.~\eqref{eq:H_QD-K_eff} into $\tau$ to avoid confusions. An important feature of Eq.~\eqref{eq:H_QD-K-QD_eff} is that the coupling between the right dot and the Kitaev wire only involves the Majorana fermions  $\hat{\mu}$ and $\hat{\gamma}_R$. In particular, setting $\eta = 0$ automatically localizes a MZM on $\hat \nu$.

We now examine the trijunction pictured in Fig.~\ref{fig:braiding_mill_inner_dots}, panel~a). It is built by tunnel-coupling the inner QDs of three 
QD-wire-QD devices, as described by the Hamiltonian: 
\begin{equation}
 \hat H_c =   t_2 \left( e^{i \phi_{12}} \hat c_1^{\dagger} \hat c_2 +  e^{i \phi_{23}}  \hat c_2^{\dagger} \hat c_3 +  e^{i \phi_{31}}  \hat c_3^{\dagger} \hat c_1+ \text{h.c.} \right);
\end{equation}
the phases depend on the microscopic details of the contacts to the Kitaev wires, and they can be adjusted with a magnetic flux threading the inner QDs. For simplicity, the absolute values of the tunnel amplitudes are identical. 

We begin our discussion by focusing on the inner part of the trijunction:
  \begin{equation}
      \hat H_{3 \text{QD}} = i \sum_{j=1}^3 \eta_j \hat \mu_j \hat \nu_j + \hat H_c .
  \end{equation}
where $\hat c_j = \frac{1}{2}(\hat \mu_j + i \hat \nu_j)$.  For $\phi_{12} = \phi_{23} = \phi_{31} = \pi/6$ and $\eta_j = -\sqrt{3} t_2$, the model hosts a fermionic zero-mode $\frac{1}{\sqrt 3}\sum_j \hat c_j$  corresponding to two MZMs; one of them is simply:
\begin{equation}\label{MZMgamma}
  \hat{\gamma}_0 = \frac{1}{\sqrt{3}} \sum_{j=1}^3  \hat \nu_j.
\end{equation}
Importantly for the following discussion, it only contains the Majorana operators $\hat \nu_j$. For generic values of the phases $\phi_{ij}$ and for absolute values of the tunnel amplitudes that are not identical, one can still tune the energy of the inner QDs so that a MZM is localized only on operators $\hat \nu_j$, but the energies of the three QDs $\eta_j$ will need to be tuned to different values, dubbed $\eta_{j,0}$.
However, if $\phi_{12}+\phi_{23}+\phi_{31} = 0 \;\rm{mod}(\pi)$ an undesired degeneracy appears, and if any of the phases $\phi_{ij}$ vanishes, $\eta_{j,0} \to \infty$. These two latter situations should be carefully avoided in an experiment; the following discussion is generically valid for all other cases.

The low-energy description of the full trijunction involves $18$ Majorana fermions; the Hamiltonian, sketched in Fig.~\ref{fig:braiding_mill_inner_dots}, panel~b), reads
\begin{equation}\label{eq:H_mill2_MF}
    \hat H_\text{3-junct.}^{\text{eff}} = \sum_{j=1}^3 \hat H_{\text{QD-K-QD}}^{\text{eff}\;(j)} + \hat H_c.
\end{equation}
For simplicity, the couplings in the different wires are chosen to be identical except for the tunable QD energies $\xi_j$. Interestingly, at $\eta_j = \eta_{j,0}$, $\hat{\gamma}_0$ in Eq.~\eqref{MZMgamma} is not coupled to the Kitaev wires by the hopping $\tau$ and thus remains a MZM localized within the three inner QDs.
We conclude that
for  $\xi_j = 0$ and $\eta_j = \eta_{j,0}$, the MZMs $\hat \alpha_{1}$, $\hat \alpha_2$, $\hat \alpha_3$ and $\hat \gamma_0$ decouple from the rest. 
The entire system can be thought of as four MZMs $\{\hat \alpha_j, \hat \gamma_0\}$ connected to a network of $14$ Majorana fermions by tunable coupling amplitudes $\{\xi_j, \eta_j\}$, as sketched in Fig.~\ref{fig:braiding_mill_inner_dots}, panel~c). 

Note that the derivation of this simple picture only requires that all energies involved in the effective Hamiltonian~\eqref{eq:H_mill2_MF} are small with respect to the superconducting gap of the wires; that is $\varepsilon, \tau, t_2, \eta_j, \xi_j \ll |\Delta|$. These conditions are convenient for discussing the braiding procedure presented in Sec.~\ref{sect:braiding_protocol}, but they can be further released: if $\tau$ is increased, the MZMs start leaking into the wires and are no longer strictly localized on QDs; however, the system still features exact and (exponentially) localized MZMs. For large values of $\xi_j, \eta_j$, the QDs are anyway effectively decoupled from the modes $\hat \alpha_j, \hat \gamma_0$, regardless of their mixing with the bulk states of the wire. In the recent experiments performed by Deng \textit{et al.}\cite{Deng_2016, Deng_2017} on a QD coupled to a Kitaev wire, $|\Delta| \sim 200 $ $\mu $eV, and $|\tau|\sim100 $ $\mu $eV (extracted from experimental data of Ref.~\onlinecite{Deng_2017}). The \textit{bowtie} pattern of Fig.~\ref{fig:eff_model_patterns} has been reproduced experimentally with $|\varepsilon|$ appearing to be of the order of the $\mu $eV.

\subsection{Braiding protocol}\label{sect:braiding_protocol}

\begin{figure}
    \centering
    \includegraphics[width = \columnwidth]{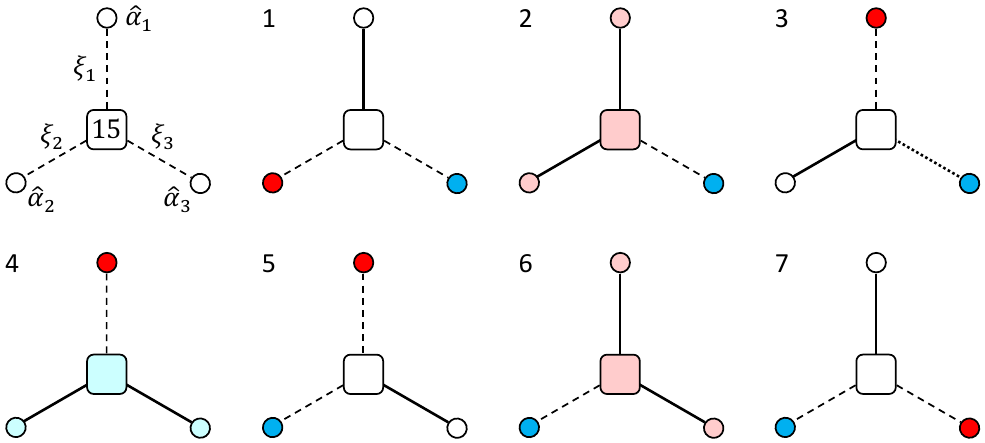}
    \caption{Braiding procedure exchanging the positions of the MZMs $\hat\alpha_2$ (red) and $\hat\alpha_3$ (blue). Couplings $\xi_j$ are switched between $\xi_\text{max}$ (coupling \textit{on}, solid line) and $0$ (coupling \textit{off}, dashed line) by tuning the QD energies. The entire protocol is based on three replicas of the transport scheme of Fig.~\ref{fig:transport}, right panel. During the entire procedure, the ground state is twofold degenerate. The QD energies $\xi_j$ in steps $1\hdots7$ are summarized in Table \ref{tab:braiding_protocol}.}
    \label{fig:braiding_protocol}
\end{figure}

As depicted in Fig.~\ref{fig:braiding_mill_inner_dots} (c), we identified for the trijunction four MZM under the fine-tuned condition $\xi_j = 0$, $\eta_j = \eta_{j,0}$. 
We now  show that by moving away from this point, we can implement transport of Majorana fermions following the principles laid out in  Sec.~\ref{sect:transport}. We discuss in particular a protocol braiding $\hat \alpha_2$ and $\hat \alpha_3$ by manipulating the QD energies. We tune $\eta_j \ne \eta_{j,0}$ during the whole protocol: the system, illustrated in Fig.~\ref{fig:braiding_protocol}, is composed of $3$ MZMs  $\hat \alpha_j$ connected to a network of $15$ Majorana fermions by three tunable couplings $\xi_j$. At the initial time ${t=0}$, we set $\xi_2 = \xi_3 = 0$, and $\xi_1 = \xi_\text{max} \neq 0$. The precise value of  $\xi_\text{max}$ is not important as long as it satisfies the condition $\xi_\text{max} \ll \Delta$. The modes $\hat\alpha_2$ and $\hat\alpha_3$ host clearly localized MZMs and the ground state is doubly degenerate.

The braiding protocol follows the one proposed in Ref.~\onlinecite{Heck_Beenakker_2012} and consists in moving across the system one MZM at each step, following the scheme explained in Fig.~\ref{fig:transport}, right panel. It requires $7$ steps, which are summarized in Table~\ref{tab:braiding_protocol} and illustrated in Fig.~\ref{fig:braiding_protocol}: $\hat \alpha_2$ is moved to position $1$ (where $\hat\alpha_1$ was), then $\hat \alpha_3$ to position $2$  and finally $\hat \alpha_2$ to position $3$. The overall effect is to interchange $\hat\alpha_2$ and $\hat\alpha_3$. $\hat \alpha_1$ plays no direct role in the exchange except for the fact that its position is used as a storage buffer for $\hat \alpha_2$.
During the entire procedure, the ground state is twofold degenerate, and higher energy levels are gapped by an energy $E_G'\ge\mathcal{O}\left(\text{min}(\tau, t_2, \varepsilon, \xi_\text{max})\right)$ which fixes a timescale for adiabaticity. Experimentally, $\varepsilon$ appears to be the smallest parameter~\cite{Deng_2016, Deng_2017}; for $\varepsilon\sim 1$ $\mu $eV transport experiments shall be performed at frequency well below the GHz regime.

\begin{table}
    \centering
    $$
    \begin{array}{c|c c c | c c c}
        \quad \text{Step} \quad & \mspace{9mu} \xi'_1 \mspace{9mu} & \mspace{9mu} \xi'_2 \mspace{9mu} & \mspace{9mu} \xi'_3\mspace{9mu} & \mspace{9mu} \hat\alpha_1(t) \mspace{9mu} & \mspace{9mu} \hat\alpha_2(t) \mspace{9mu} & \mspace{9mu} \hat\alpha_3(t) \mspace{9mu}\\
        \hline
        1 & \xi_\text{max} & 0 & 0 & * & \hat\alpha_2 & \hat\alpha_3 \\
        2 & \xi_\text{max} & \xi_\text{max} & 0 & * & * & \hat\alpha_3\\
        3 & 0 & \xi_\text{max} & 0 & -\hat\alpha_2 & * & \hat\alpha_3\\
        4 & 0 & \xi_\text{max} & \xi_\text{max} & -\hat\alpha_2 & * & * \\
        5 & 0 & 0 & \xi_\text{max} & -\hat\alpha_2 & -\hat\alpha_3 & * \\
        6 & \xi_\text{max} & 0 & \xi_\text{max} & * & -\hat\alpha_3 & * \\
        7 & \xi_\text{max} & 0 & 0 & * & -\hat\alpha_3 & \hat\alpha_2
    \end{array}
    $$
    \caption{QD energies  $\xi_j$ at each step of the braiding protocol described in Fig.~\ref{fig:braiding_protocol}, and tracking of the initial MZMs $\hat\alpha_i$ through the transport steps. A star ($*$) indicates that $\hat \alpha_i(t)$ does not correspond to a MZM. The $C_3$ symmetry of the trijunction imposes $\chi_1 = \chi_2$ in Eq.~\eqref{eq:braiding_sign}, and provides a simple way of keeping track of the signs.}
    \label{tab:braiding_protocol}
\end{table}
The full braiding operation after time $T$ transfers the MZMs as
\begin{equation}\label{eq:braiding_outcome}
    \hat\alpha_2(T) = - \zeta \hat \alpha_3 \qquad \hat\alpha_3(T) =  \zeta \hat \alpha_2,
\end{equation}
where we have written $\hat \alpha_j$ for $\hat \alpha_j(0)$, and where $\zeta = \pm 1$ is the chirality of the braiding\cite{Clarke_Sau_2011}; in Fig.~\ref{fig:braiding_protocol} and Table~\ref{tab:braiding_protocol}, $\zeta =1$. The associated unitary time-evolution operator is~\cite{Ivanov_2001}:
\begin{equation}\label{eq:braiding_operator}
    \hat U_{23} = \text{exp}\left(\frac{\pi}{4} \zeta \hat \alpha_2 \hat \alpha_3\right) = \frac{1}{\sqrt{2}}\left( 1 + \zeta \hat \alpha_2 \hat \alpha_3\right),
\end{equation}
so that $\hat\alpha_j(T) = \hat U_{23}^\dagger \hat\alpha_j \hat U_{23}$.

Repeating the braiding twice results in a non-trivial operation on the degenerate ground state:
\begin{equation}\label{eq:double_braiding_outcome}
    \left(\hat U_{23}\right)^2 = \zeta \hat \alpha_2 \hat \alpha_3.
\end{equation}
This is the signature of the non-Abelian statistics of the MZMs. One recovers the initial state (up to a global phase) only after performing the braiding $4$ times.

\subsection{Experimental demonstration of non-Abelian statistics}\label{sect:experimental_demo}

The previous setup allows to braid two MZMs, but the time-evolution operator $\hat U_{23}$ in Eq.~\eqref{eq:braiding_operator} commutes with the parity $\hat P_{23} = i\hat \alpha_2 \hat \alpha_3$. This limits the operations that can be performed on the twofold degenerate ground space through braiding to a dephasing between the even-parity and odd-parity states, which cannot be measured experimentally. 
This issue can be circumvented by increasing the number of MZMs so that for fixed parity the ground space is degenerate\cite{DasSarma_2015}; in the following, we propose and discuss a setup hosting six MZMs.

\begin{figure}
    \centering
    \includegraphics[width = .9\columnwidth]{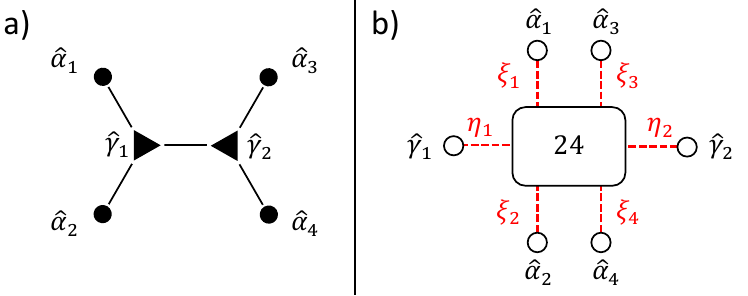}
    \caption{a) Circuit with $5$ Kitaev wires (solid lines), and two trijunctions (triangles) hosting six MZMs. The outer QDs (circles) allow for parity measurements.
    b) The circuit breaks down to $6$ MZMs $\{\hat \alpha_{1\hdots4}, \hat \gamma_{1,2}\}$ connected to a network of $24$ Majorana fermions by tunable couplings $\xi_{j}, \eta_j$.
    }
    \label{fig:1Qubit}
\end{figure}

We assemble $5$ Kitaev wires so that they form two trijunctions as shown in Fig.~\ref{fig:1Qubit}, panel~a). 
With arguments similar to those presented in Sec.~\ref{sect:trijunction}, we obtain that the subgap physics is composed of $6$ MZMs $\{\hat \alpha_{1}, \hat \alpha_{2}, \hat \alpha_{3}, \hat \alpha_{4}, \hat \gamma_1,\hat \gamma_2\}$ connected to a network of $24$ Majorana fermions by tunable couplings $\xi_{j}, \eta_j$\footnote{MZMs $\hat \gamma_1$ and $\hat \gamma_2$ are actually connected to the network through three couplings that could be called $\eta_{j, 1}$ and $\eta_{j, 2}$, $j=1\hdots3$, similarly to the simpler case shown in Fig.~\ref{fig:braiding_mill_inner_dots}, panel c). For simplicity, we only denote them as $\eta_1$ and $\eta_2$, the relevant question being whether the couplings are \textit{on} or \textit{off} and not the number of links} (see Fig.~\ref{fig:1Qubit}, panel b).
At the beginning two couplings are set out of resonance, for instance $\xi_3 = \xi_4 = \xi_{\rm max}$. The four remaining MZMs are strictly localized on $\{\hat \alpha_1, \hat \alpha_2, \hat \gamma_1, \hat \gamma_2 \}$, and the ground space is fourfold degenerate.
If we define the total parity $\hat P_{1122} = - \hat \alpha_1 \hat \alpha_2 \hat \gamma_1 \hat \gamma_2 $, 
two ground states have even parity and two have odd parity. 

\begin{figure}
    \centering
    \includegraphics[width = \columnwidth]{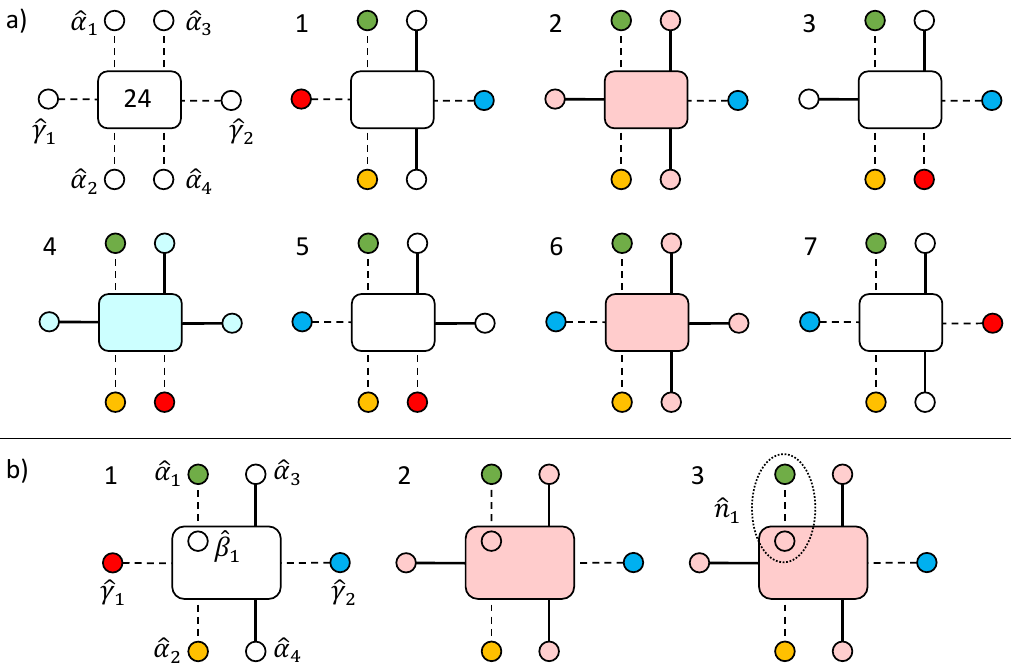}
    \caption{a) Braiding of $\hat \gamma_1$ (red) and $\hat \gamma_2$ (blue). The couplings are modified in time, similarly to the protocol presented in Sec.~\ref{sect:braiding_protocol}; dashed lines indicate that a coupling is \textit{off} (QD energy at resonance), and solid lines stand for couplings \textit{on} (QD energy out of resonance).
    b) Charge measurement used to infer the parity $\hat P_{11} = i \hat \alpha_1 \hat \gamma_1$. $\eta_1$ is set out of resonance, so that the MZM initially localized on $\hat \gamma_1$ (red) expands over the entire network. In particular, it acquires a nonzero component on $\hat \beta_1$, which is the second Majorana mode of the QD hosting $\hat \alpha_1$ (green). The charge measured on this QD is correlated to the parity $\hat P_{11}$, as described in Eq.~\eqref{eq:expectation_charge_qd}.}
    \label{fig:experimental_demo}
\end{figure}

In Fig.~\ref{fig:experimental_demo}, panel a), we show how to braid $\hat \gamma_1$ and $\hat \gamma_2$ by only tuning QD energies; even if the time-evolution operator $\exp\left[ \frac{\pi}{4} \zeta \hat \gamma_1 \hat \gamma_2 \right]$ commutes with $\hat P_{1122}$, the non-Abelian statistics can be experimentally demonstrated.
The idea is to measure the parity $\hat P_{11} = i \hat \alpha_1 \hat \gamma_1$  before and after performing two consecutive braidings of $\hat \gamma_1$ and $\hat \gamma_2$. These two consecutive braidings amount to the operation ${\hat \gamma_{1,2} \rightarrow -\hat \gamma_{1,2}}$ as in Eq.~\eqref{eq:double_braiding_outcome}, and therefore imply ${\hat P_{11} \rightarrow -\hat P_{11}}$. Because the initial and the final Hamiltonians are identical, this parity switch is a signature of the non-Abelian nature of MZMs\cite{Clarke_Sau_Jay_2017}.

We now outline a simple protocol that gives some information about the parity $\hat P_{11}$; it is sketched in Fig~\ref{fig:experimental_demo}, panel b): starting from the initial situation with $\xi_3 = \xi_4 = \xi_{\rm max}$, we adiabatically tune the energy $\eta_{j} $ of the three quantum dots in the first trijunction to the out-of-resonance value $ \eta_{\rm max}$.  As a consequence, the MZM initially localized on $\hat \gamma_1$ spreads over the entire network. We call $\hat \gamma_1'$ this delocalized MZM and, because of adiabaticity, $\hat P_{11} = \hat P_{11}'$ where $ \hat P_{11}' = i \hat \alpha_1 \hat \gamma_1' $. In particular,
$\hat \gamma_1'$ acquires a nonzero component $u$ on $\hat \beta_1$, but no component on the decoupled MZM $\hat \alpha_1$.
As such, the measurement of the occupation of the QD~$1$: 
\begin{equation}
    \hat n_1 = \frac{1}{2} \left(1 + i \hat \alpha_1\hat\beta_1\right),
\end{equation}
can distinguish between two parity states of $\hat P_{11}$ with an accuracy fixed by $u$. More precisely, if the system is in an eigenstate of $\hat P_{11}$ with parity $p$, the expectation value of $\hat n_1$ is (see Appendix~\ref{app:simple_measurement}):
\begin{equation}\label{eq:expectation_charge_qd}
    \braket{\hat n_1}_p = \frac{1}{2}(1+up).
\end{equation}

Many experimental techniques allow for this charge measurement; for instance, it can be performed with a quantum point contact placed nearby, or through microwave reflection\cite{Wiel_Franceschi_2002, Hanson_Kouwenhoven_2007, Hu_Churchill_2007, Barthel_Reilly_2009, Persson_Wilson_2010, Petersson_Smith_2010, Hu_Kuemmeth_2011, Jung_Schroer_2012, Medford_Beil_2013, Colless_Mahoney_2013, Eng_Ladd_2015, Stehlik_Liu_2015}. Because $|u|<1$, this measurement does not correspond to an exact readout of the parity $\hat P_{11}$, and the final state is projected into a tilted basis (the charge eigenstates instead of the parity eigenstates). Therefore, the non-Abelian nature of the MZMs should be deduced by accumulating statistics on the outcome of the two charge measurements (before and after braiding). In a circuit with uniform tunnel couplings $\tau$, $u$ saturates at $1/\sqrt{5}$, but it can be further increased by releasing this constraint (see Appendix~\ref{app:accuracy}).
Alternatively, one can use the more sophisticated measurement scheme presented in Appendix~\ref{app:strong_measurement}, which allows for an exact measurement of $\hat P_{11}$. Note that these measurement schemes can also read out parities of type $i\hat\alpha_j\hat\alpha_k$, and that they recycle the QDs originally used for braiding into parity meters.

\section{Topological quantum computation in QD-controlled circuits}\label{sec:tqc}

The setup presented in Sec.~\ref{sect:experimental_demo} and sketched in Fig.~\ref{fig:1Qubit} provides two zero-energy states for a fixed parity, which can be used as a logical qubit. It is encoded by four MZMs, renamed $\hat \gamma_{1\hdots4}$ for simplicity, that are spatially separated in the circuit: quantum information is stored non-locally and therefore enjoys topological protection. A representation of the logical Pauli matrices on both parity sectors is given by:
\begin{equation}\label{eq:logical_pauli}
    \bar \sigma_x = -i\hat \gamma_1 \hat \gamma_2 \qquad 
    \bar \sigma_y = -i\hat \gamma_2 \hat \gamma_3 \qquad 
    \bar \sigma_z = -i\hat \gamma_1 \hat \gamma_3 ,
\end{equation}
and we define the logical states $\ket{\bar 0}$ and $\ket{\bar 1}$ as the eigenvectors of $\bar \sigma_z$.

Manipulations of the MZMs perform quantum logical operations, and 
the braiding of MZMs $\hat \gamma_i$ and $\hat \gamma_j$ with chirality $\zeta=\pm1$ according to the protocol shown in Fig~\ref{fig:experimental_demo}, panel a) is described by the time-evolution operator:
\begin{equation}\label{eq:braiding_gate}
    \hat U_{ij} = \text{exp}\left(\frac{\pi}{4} \zeta \hat \gamma_i \hat \gamma_j\right).
\end{equation}
Braiding operations therefore implement discrete single-qubit $\pi/2$ rotations along the $x$, $y$ and $z$ axis of the Bloch sphere defined by Eq.~\eqref{eq:logical_pauli}. A standard choice for reaching universality is to complete them with the single-qubit $\pi/8$ phase gate~$T$ and the two-qubits controlled~$\sigma_z$ gate~$\Lambda(\sigma_z)$. 

An implementation of $\Lambda(\sigma_z)$ with QD-controlled circuits is given in Sec.~\ref{sect:Cz_gate}; it is achieved using braiding operations together with projective measurements of $4$-MZMs parities.
The implementation of the $T$~gate is discussed in Sec.~\ref{sect:T_gate}, and relies on magic state distillation instead of topological protection. 

\subsection{\texorpdfstring{$4$-MZMs parity measurement and $\Lambda(\sigma_z)$ gate}{}}\label{sect:Cz_gate}

Let us consider two replicas $a$ and $b$ of our single-qubit circuit. Both of them operate four MZMs, $\hat \gamma_{1\hdots4}^a$ and $\hat \gamma_{1\hdots4}^b$ respectively, which encode the topological qubits $\ket{q_a}$ and $\ket{q_b}$. Despite MZMs from different circuits cannot be braided, it has been shown that a controlled~$\sigma_z$ gate over $\ket{q_a}\otimes\ket{q_b}$ could be implemented only through braiding and projective parity measurements\cite{Bravyi_kitaev_2002, Bravyi_2006}. Braiding and parity readout are the natural operations one can perform with MZMs, and together with the $\Lambda(\sigma_z)$, they implement the set of gates known as the Clifford group. The exact scheme for a $\Lambda(\sigma_z)$ gate based only on these natural operations is shown in Appendix~\ref{app:Cz_gate}; here we only discuss its feasibility in QD-controlled circuits. In addition to on-circuit braiding and on-circuit parity readout (implemented in Sec.~\ref{sect:experimental_demo}), the procedure requires:
\begin{enumerate}[(i)]
    \item An ancillary pair of MZMs lying on the physical circuit of the target qubit.
    \item Projective measurement of a 4-MZMs parity of type:
    \begin{equation}
        \hat P_{jklm} = -\hat \gamma_j^a \hat \gamma_k^a \hat \gamma_l^b \hat \gamma_m^b.
    \end{equation}
\end{enumerate}

\begin{figure}
    \centering
    \includegraphics[width = .8\columnwidth]{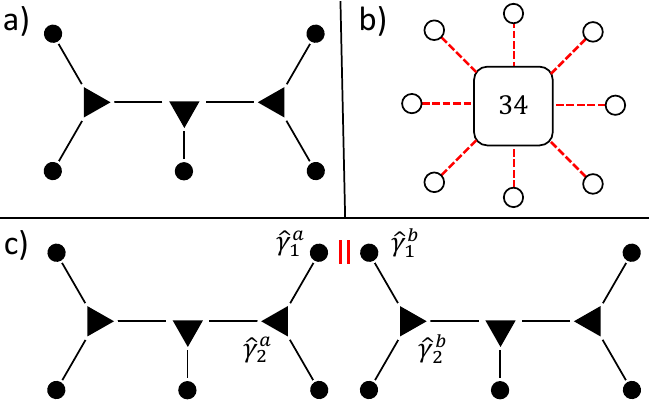}
    \caption{a) 2-qubits circuit allowing for the implementation of a $\Lambda(\sigma_z)$ gate.
    b) It amounts to $8$ Majorana fermions coupled to a network; $2$ are gapped and serve for transport purposes, and the $6$ remaining ones store two qubits: a logical qubit and an ancilla.
    c) $4$-MZMs parity measurement between two 2-qubits circuits. The simple parities $\hat P^a_{12} = i \hat \gamma^a_1 \hat \gamma^a_2$ and $\hat P^b_{12} = i \hat \gamma^b_1 \hat \gamma^b_2$ can be perfectly correlated with the charges $q_a$ and $q_b$ on the neighbouring QDs. A parity meter (red bars~$||$) measuring the parity of the total charge $Q=q_a+q_b$ performs a  projective measurement of the $4$-MZMs parity $\hat P_{1212} = -\hat \gamma_1^a \hat \gamma_2^a \hat \gamma_1^b \hat \gamma_2^b$.
    }
    \label{fig:cross_parity}
\end{figure}

The ancillary pair of MZMs can be provided by increasing the size of our elementary circuit: a version involving three trijunctions and $8$ tunable Majorana fermions, therefore providing $6$ computational MZMs and hosting two qubits (later called \textit{2-qubits circuit}), is sketched in Fig.~\ref{fig:cross_parity}, panels~b) and~c). For simplicity, we will not distinguish between the control and the target qubit, and we will store both of them on such a scaled-up circuit. The computational MZMs are from now on $\hat \gamma_{1\hdots6}^a$ and $\hat \gamma_{1\hdots6}^b$. Braiding and parity readout schemes inside these circuits is analogous to those of Fig.~\ref{fig:experimental_demo}.

In Sec.~\ref{sect:experimental_demo} and Appendix~\ref{app:parity_readout}, we have shown how the parities $\hat P^a_{jk} = i \hat \gamma^a_j \hat \gamma^a_k$ and $\hat P^b_{lm} = i \hat \gamma^b_l \hat \gamma^b_m$ could be perfectly correlated with the occupation number of QD~$a_j$ (hosting~$\hat \gamma^a_j$) and QD~$b_l$ (hosting~$\hat \gamma^b_l$) respectively. Therefore, measuring the total charge parity of the system made of QDs $a_j$ and $b_l$ amounts to a projective measurement of $\hat P^a_{jk}\hat P^b_{lm} =  \hat P_{jklm}$. This can be considered experimentally if these two QDs are placed nearby (see Fig.~\ref{fig:cross_parity}, panel a): there have been proposals for building charge parity meters, either with a quantum point contact placed between the two adjacent QDs\cite{Trauzettel_Jordan_2006}, or through microwave reflection\cite{Schroer_2012}. 

\subsection{\texorpdfstring{Magic state distillation and $\pi/8$ phase gate}{}}\label{sect:T_gate}

A direct consequence of the Gottesman-Knill theorem~\cite{Gottesman_1998} is that operations of the Clifford group are not sufficient to implement the $T$ quantum gate. One could of course use unprotected operations in order to implement it, but the resulting errors would spoil the benefits of topological quantum computation. For example, splitting the two logical states of a qubit in energy during a precise period could amount to a $\pi/4$ relative phase, and this could be done in a QD-controlled circuit by coupling two computational MZMs.

An efficient implementation of the $T$ gate can still be achieved. It relies on error correcting codes, and is known \textit{magic state distillation}~\cite{Bravyi_kitaev_2005, Bravyi_2006}. Let us assume that we want to operate the $T$ gate on a logical qubit $\ket {q_1}$, and that we have access to a second qubit prepared in the \textit{magic state}: 
\begin{equation}
    \ket H = \frac{1}{\sqrt{2}}\left(\ket{0} + e^{i \pi/4}\ket{1} \right).
\end{equation}
Then, applying a proper set of Clifford gates on $\ket{H}\otimes\ket{q_1}$ (implemented in Sec.~\ref{sect:Cz_gate}) amounts to performing a $T$ gate on $\ket{q_1}$~\cite{DasSarma_2015}.

Of course, a magic state cannot be prepared with the Clifford group, but it can be approached with good accuracy by \textit{distillation protocols}\cite{Bravyi_kitaev_2005, Bravyi_2006, Bravyi_Haah_2012, Meier_Eastin_2012,  Paetznick_Reichardt_2013, Jones_2013, Eastin_2013}. Such schemes require several noisy copies $\{\ket{H'_i}\}$ of $\ket{H}$, that can be obtained through unprotected operations. A distillation step consists in measuring a set of stabilizers on $\{\ket{H'_i}\}$, and in implementing some corresponding corrections; if the initial error on  $\{\ket{H'_i}\}$ are small enough, the protocol builds a converging copy of $\ket{H}$. Importantly, the distillation routine only relies on the Clifford group, and can therefore be implemented in QD-controlled circuits. In a topological quantum processor, distillation protocols should run continuously in dedicated registers, so that magic states are always available when needed for a logical operation.

\subsection{Scaling up}

\begin{figure}
    \centering
    \includegraphics[width = .8\columnwidth]{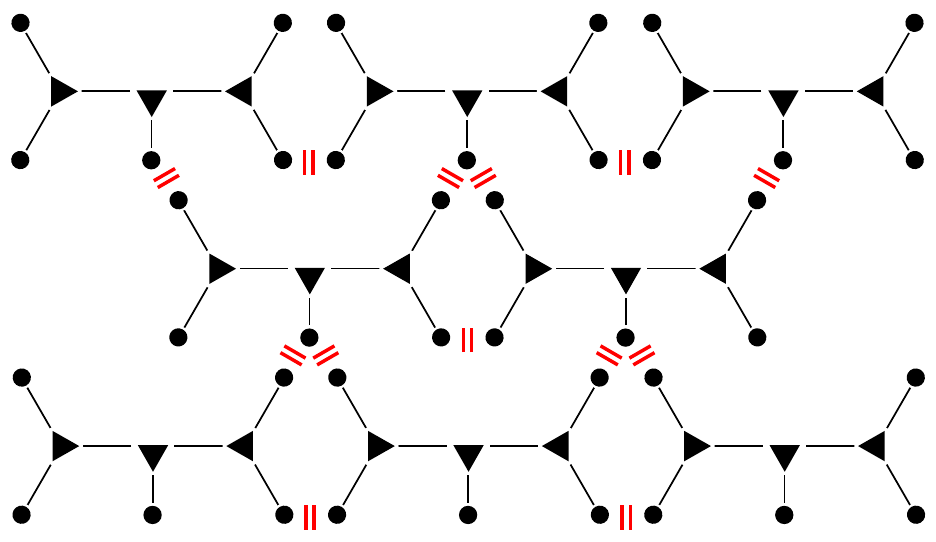}
    \caption{Network of $8$ replicas of the 2-qubits circuit (one logical qubit and one ancilla) presented in Sec.~\ref{sect:Cz_gate}, virtually connected by QD charge parity meters (red bars $||$). This device amounts to a 8-qubit topological quantum processor, and the underlying pattern can be scaled-up at will.
    }
    \label{fig:scale_up}
\end{figure}

A straight-forward way of scaling up a QD-controlled topological quantum processor is to build a network of 2-qubits circuits, each of them hosting a single logical qubit and one ancilla (needed for the controlled $\sigma_z$ operation). As such, two-qubit operations can be performed on neighbouring sites through $4$-MZMs parity measurements, and quantum information is encoded in a sparse way, which prevents local errors from propagating. An example of such architecture, which can virtually be extended at will, is given in Fig.~\ref{fig:scale_up}. 

\section{Conclusions}
\label{sect:conclusion}

Motivated by the recent experimental realization of a nanowire hosting MZMs coupled to a single QD~\cite{Deng_2016, Deng_2017}, we discussed a network of Kitaev's wires coupled to tunable QDs where topologically-protected operations can be achieved by solely manipulating the QDs.
In particular, we described in details a trijunction with six QDs where the non-Abelian statistics of the MZMs can be revealed, and outlined ideas for scaling up the network to more advanced purposes.
These results show the exceptional versatility and usefulness of hybrid nanowire-QD devices that are currently produced in laboratories.

\acknowledgments

We thank M. Burrello and A. Stern for enlightening observations and comments; we also gratefully acknowledge discussions with N. Clausen, V. Crepel and J. Lebreuilly.
L. Mazza was  supported  by  LabEX  ENS-ICFP:  ANR-10-LABX-0010/ANR-10-IDEX-0001-02 PSL*.  
C. Mora acknowledges support  from  Idex  PSL  Research  University  (ANR-10-
IDEX-0001-02 PSL)

\appendix

\section{Parity switch}\label{app:parity_switch}

The effective model for the QD-wire setup is in Eq.~\eqref{eq:H_QD-K_eff}, which we reproduce here below for better readability:
\begin{equation}
    \hat H_{\text{QD-K}}^{\text{eff}} = i \left( \xi \,\hat{\alpha} \hat{\beta} + \varepsilon \, \hat{\gamma}_L\hat{\gamma}_R + \tau_L \, \hat{\beta} \hat{\gamma}_L + \tau_R \hat{\alpha}\hat{\gamma}_R \right).
\end{equation}
It is a quadratic form of the vector $\{\hat \alpha$, $\hat \beta$, $\hat{\gamma}_L$, $\hat{\gamma}_R\}$:
\begin{equation}
    \hat H_{\text{QD-K}}^{\text{eff}} = \frac{i}{2} \sum_{i,j} x_i A_{ij} x_j , \qquad \vec{x}^T = 
    \left(
    \begin{array}{cccc}
        \hat \alpha & \hat \beta & \hat \gamma_L & \hat \gamma_R
    \end{array}
    \right);
\end{equation}
where:
\begin{equation}
    A_{i,j} = 
    \left(
    \begin{array}{cccc}
        0 & \xi & 0 & \tau_R \\
        -\xi & 0 & \tau_L & 0  \\
        0 & - \tau_L & 0 & \varepsilon  \\
        -\tau_R & 0 & - \varepsilon & 0
    \end{array}
    \right).
\end{equation}
The condition for $\hat H_{\text{QD-K}}^{\text{eff}}$ to exhibit MZMs is:
\begin{equation}
    0 = \text{Det}\; A_{ij} = \left( \varepsilon \xi + \tau_L \tau_R \right)^2;
\end{equation}
and we recover the position of the parity switch given in Eq.~\eqref{eq:resonance}.

\section{Parity readout}\label{app:parity_readout}

\subsection{Simple readout protocol}\label{app:simple_measurement}

The readout protocol proposed in Sec.~\ref{sect:experimental_demo} aims at measuring the parity $\hat P_{11}(0) = i \hat \alpha_1(0) \hat \gamma_1(0)$. During a first step, the couplings $\eta_j$ of the first trijunction are tuned out of resonance while the MZMs $\{\hat \alpha_1, \hat \alpha_2, \hat \gamma_2\}$ are kept uncoupled, so that after a time $T_1$:
\begin{align}\label{eq:mzms_readout}
    \nonumber
    &\hat \alpha_1 (0) = \hat \alpha_1 (T) \qquad \hat \alpha_2 (0) = \hat \alpha_2 (T) \qquad \hat \gamma_2(0) = \hat \gamma_2 (T) \\
    &\hat\gamma_1(0) = u\hat \beta_1(T) + \sum_j v_j \hat \delta_j(T),
\end{align}
where the set $\{\hat\delta_j(T)\}$ stands for all Majorana fermions of the effective model, except for $\{\hat \alpha_1(T), \hat \alpha_2(T), \hat \gamma_2(T)\}$ and $\hat \beta_1(T)$. Therefore:
\begin{align}\label{eq:parity_decomposition}
    \hat P_{11}(0) &= i \hat \alpha_1 (0) \hat \gamma_1 (0) \\
    \nonumber
    &= i u \hat \alpha_1 (T)\hat \beta_1(T) + \sum_j i v_j \hat \alpha_1 (T) \hat \delta_j(T)
\end{align}

We now measure the occupation of QD~$1$. Its expectation value on a state $\ket \psi$ is: 
\begin{equation}
    \bra{\psi}\hat n_1(T)\ket{\psi} = \frac{1}{2} \left(1 + \bra{\psi}i \hat \alpha_1(T)\hat\beta_1(T)\ket{\psi}\right).
\end{equation}
If the system is in an eigenstate $\ket{\psi_p}$ of $\hat P_{11}(0)$ with parity $p=\pm 1$, we can enforce the equality:
\begin{align}
    \bra{\psi_p}i \hat \alpha_1(T)&\hat\beta_1(T)\ket{\psi_p}= \\
    \nonumber
    &= \frac{p}{2}\bra{\psi_p}\{i \hat \alpha_1(T)\hat\beta_1(T), \hat P_{11}(0)\}\ket{\psi_p}.
\end{align}
In the decomposition \eqref{eq:parity_decomposition}, all terms of $\hat P_{11}(0)$, except for the first one, anticommute with $i \hat \alpha_1(T)\hat\beta_1(T)$. Therefore:
\begin{align}
    \bra{\psi_p}&i \hat \alpha_1(T)\hat\beta_1(T)\ket{\psi_p}= \\
    \nonumber
    &= \frac{p}{2}\bra{\psi_p}\{i \hat \alpha_1(T)\hat\beta_1(T), i u \hat \alpha_1(T)\hat\beta_1(T)\}\ket{\psi_p}\\
    \nonumber
    &= up,
\end{align}
and as stated in Sec.~\ref{sect:experimental_demo}:
\begin{equation}\label{eq:expectation_charge_qd_app}
    \bra{\psi_p}\hat n_1(T)\ket{\psi_p} = \frac{1}{2}\left( 1+up\right)
\end{equation}

Note that the MZMs $\hat \alpha_2$ and $\hat \gamma_2$, which remain uncoupled during the entire procedure and upon which no measurement is performed, are not affected by this measurement.

\subsection{Accuracy}\label{app:accuracy}

The accuracy of the previous protocol is set by the weight $u$ of the delocalized MZM in Eq.~\eqref{eq:mzms_readout}, where $|u|=1$ corresponds to a perfect measurement and $|u|=0$ gives no information about the parity (see Eq.~\eqref{eq:expectation_charge_qd_app}). We have previously set $\xi_3 = \xi_4 = \xi_\text{max}$ and $\eta_1 = \eta_\text{max}$ (situation 2 of Fig.~\ref{fig:experimental_demo}, panel b); in the limit $\xi_\text{max}, \eta_\text{max}\rightarrow \infty$, we find the simple expression:
\begin{equation}
    |u| = \frac{|r|}{\sqrt{5 r^2+5}}, \qquad r = \frac{\varepsilon}{\tau},
\end{equation}
which is maximal for a weakly coupled QD ($r^2\ll1$) and saturates at $1/\sqrt{5}\approx0.45$. 

This factor actually arises from the delocalized MZM spreading uniformly among the resonant QDs (one hosting $\hat \alpha_1$, one hosting $\hat\alpha_2$ and three hosting $\hat\gamma_2$). In principle, this spreading could be biased by taking non-uniform physical parameters for the circuit. More precisely, if $\tau_{1,2}$ are the tunnel couplings of the outer QDs hosting $\hat\alpha_{1,2}$, $\varepsilon_{1,2}$ the energy splittings of their adjacent Kitaev wires, $\tau_0$ the tunnel coupling of the inner QDs to the wires and $\varepsilon_0$ the energy splitting of the central wire, then in the same limit $\xi_\text{max}, \eta_\text{max}\rightarrow \infty$:
\begin{equation}
    |u| = \frac{|r_1|}{\sqrt{r_1^2 + r_2^2 + 3r_0^2 +5}}, \qquad r_j = \frac{\varepsilon_j}{\tau_j},
\end{equation}
and $|u|=1$ in the limit where only the QD $1$ is weakly coupled.

\subsection{Exact parity measurement}\label{app:strong_measurement}

An intuitive idea for reading out the parity of a topological system consists in splitting the even, and odd states in energy and then performing some spectroscopic measurement. If one can turn the existing QDs of the system into a spectrometer, exact parity readout is achieved without any additional cost. Such a protocol was put forward in Ref.~\onlinecite{Gharavi_Baugh_2016}, where Rabi oscillations between a QD and a Kitaev wire are used to determine the parity of the degenerate ground state. 

Let us start with situation $2$ of Fig.~\ref{fig:experimental_demo} (b), where the MZMs of the system are described by Eq.~\eqref{eq:mzms_readout}. Adiabatically tuning $\xi_1$ out of resonance introduces a coupling between $\hat \alpha_1(T)$ and $\hat \beta_1(T)$, and therefore splits the two parity states in energy. Tuning it further away up to the limit $\xi_1 = -\infty$, still adiabatically, charges the QD~$1$ and effectively uncouples it from the rest of the system: eigenstates of the total effective model are eigenstates of $\hat n_1$. At this point, the two parity eigenstates have been split by an energy $2\varepsilon_0$, which we assume to be much smaller than the other non-zero energy states of the Hamiltonian. The energy of QD~$1$ is quickly set at $\xi_1 = -\varepsilon_0$, allowing for Rabi oscillations between the QD and the rest of the system thanks to the coupling term:
\begin{equation}
    \hat H_\tau = i \tau \hat \beta_1 \hat \gamma_{L_1} .
\end{equation}
Importantly:
\begin{enumerate}[(i)]
    \item The Hamiltonian is still twofold degenerate because of the MZMs $\hat \alpha_2$ and $\hat \gamma_2$, but as they are strictly localized onto their respective QDs, they do not get coupled by $\hat H_\tau$.
    \item The Majorana modes defining the parity eigenstate have a non-zero component $u$ on $\hat \gamma_{L_1}$, so that they get coupled to the QD~$1$.
    \item We assume that all other energy levels are separated from $\pm \varepsilon_0$ by an energy much larger than $\tau$, so that spurious subgap states are effectivelly uncoupled from the dot.
\end{enumerate}
With these conditions, the dynamics is reduced to Rabi oscillations occurring between the QD~$1$ and the parity eigenstate. In the Fock basis associated to this subspace, the Hamiltonian reads\cite{Gharavi_Baugh_2016, Clarke_2017}:
\begin{align}
    \hat H_\text{Rabi} = &
    \left(
    \begin{array}{c}
        \ket{00}\\
        \ket{11}\\
    \end{array}
    \right)^T
    \left(
        \begin{array}{cc}
        0 & u\tau/2  \\
        u\tau/2  & \varepsilon_0 +\xi_1  \\
    \end{array}
    \right)
    \left(
    \begin{array}{c}
        \bra{00}\\
        \bra{11}\\
    \end{array}
    \right)\\
    \nonumber
    & + 
    \left(
    \begin{array}{c}
        \ket{01}\\
        \ket{10}\\
    \end{array}
    \right)^T
    \left(
        \begin{array}{cc}
        \varepsilon_0  & u\tau/2 \\
        u\tau/2  & \xi_1  \\
    \end{array}
    \right)
    \left(
    \begin{array}{c}
        \bra{01}\\
        \bra{10}\\
    \end{array}
    \right).
\end{align}
Having tuned $\xi_1 = -\varepsilon_0$ sets the Rabi process ${\ket{00}\leftrightarrow\ket{11}}$ on resonance, while ${\ket{01}\leftrightarrow\ket{10}}$ is out of resonance. We assume that the QD is weakly coupled to the wire, that is $(u \tau)^2 \ll \varepsilon_0^2$, so that this second process is completely suppressed. 

Because the QD was initially charged, the system starts either in state $\ket{10}$, and its dynamics is frozen, or in state $\ket{11}$, and it undergoes full Rabi oscillations with $\ket{00}$ at a frequency $\omega_0 = u\tau/2$. Therefore, after half a Rabi oscillation, a charge measurement on the QD can perfectly distinguish between the initial parity states. At this precise moment, the QD energy is quickly tuned out of resonance: the QD charge has been perfectly correlated with the parity we need to measure, and can now be read out by charge sensing, or be used in a $4$-MZMs parity measurement (see Sec.~\ref{sect:Cz_gate}). 

The significant advantage of this method is that the accuracy no longer saturates at $1/\sqrt{5}$ in the limit of a weakly coupled QD, essentially because QD $1$ plays a privileged role. Also, the constraint of a weakly coupled QD can in principle be further released. If the Rabi process ${\ket{01}\leftrightarrow\ket{10}}$ cannot be suppressed, it will occur with an detuned frequency $\omega_1 = \sqrt{\varepsilon_0^2+(u\tau)^2}$; if the two Rabi processes are in perfect phase opposition at a time $T_1$, both parities can perfectly be distinguished. This can be ensured by adjusting $\eta_\text{max}$ so that $\omega_1/\omega_0$ becomes a rational ratio.

Note however that this parity readout operation is not protected: an error in the duration of the Rabi oscillations will unavoidably result into an tilted projection basis.

\section{\texorpdfstring{Clifford-based $\Lambda(\sigma_z)$ gate}{}}\label{app:Cz_gate}

Here we reproduce an algorithm introduced in Refs~\onlinecite{Bravyi_kitaev_2002,Bravyi_2006}. We want to implement a $\Lambda(\sigma_z)$ gate on two topological qubits $\ket{q_a}$ and $\ket{q_b}$ built on $\hat \gamma_{1\hdots4}^a$ and $\hat \gamma_{1\hdots4}^b$ respectively. If $\ket{q_a}$ is the control qubit and $\ket{q_b}$ the target, the controlled~$\sigma_z$ gate is:
\begin{equation}
    \Lambda(\sigma_z) = {\rm exp}\left(i \frac{\pi}{4} (I-\bar\sigma_z^a) (I-\bar\sigma_z^b)\right)
\end{equation}
In terms of MZMs (see Eq.~\eqref{eq:logical_pauli}): 
\begin{align}
    \Lambda(\sigma_z) = &e^{i\pi/4} {\rm exp}\left(-\frac{\pi}{4} \hat\gamma_1^a \hat\gamma_3^a\right) {\rm exp}\left(-\frac{\pi}{4} \hat\gamma_1^b \hat\gamma_3^b\right)\\
    \nonumber
    & \times {\rm exp}\left(-i\frac{\pi}{4} \hat\gamma_1^a \hat\gamma_3^a \hat\gamma_1^b \hat\gamma_3^b\right)
\end{align}

The first two operations represent on-circuit braidings (see Eq.~\eqref{eq:braiding_gate}), which have already been discussed. We focus on the implementation of the four MZM gate:
\begin{equation}
    U^{(4)} = {\rm exp}\left(-i\frac{\pi}{4} \hat\gamma_1^a \hat\gamma_3^a \hat\gamma_1^b \hat\gamma_3^b\right)
\end{equation}
We assume that an ancillary pair of MZMs $\hat \gamma_5^b$ and $\hat \gamma_6^b$ lies on circuit b, and that it has been initialized in state $\ket{0}$ so that:
\begin{equation}\label{eq:ancillary_mzm}
    (\hat \gamma_5^b+i\hat \gamma_6^b)\ket{\psi} = 0,
\end{equation}
where $\ket \psi$ is the initial wave function of the entire computational subspace.

We first measure the $4$-MZMs parity $\hat P_{1335} = - \hat\gamma_1^a \hat\gamma_3^a \hat\gamma_3^b \hat\gamma_5^b$ (demonstrated in Sec.~\ref{sect:Cz_gate}), and depending on the output $p_1 = \pm1$, the initial wave function gets projected by:
\begin{equation}
    \Pi_1^{(p_1)} = \frac{1}{2}\left( 1 - p_1\hat\gamma_1^a \hat\gamma_3^a \hat\gamma_3^b \hat\gamma_5^b\right).
\end{equation}
We then measure the on-circuit parity $\hat P_{15}^b = i\hat \gamma_1^b\hat\gamma_5^b$. An outcome $p_2 = \pm1$ amounts to the projector:
\begin{equation}
    \Pi_2^{(p_2)} = \frac{1}{2}\left( 1 + p_2\hat\gamma_1^b\hat\gamma_5^b\right).
\end{equation}
Then, depending on the measured parities $p_1$ and $p_2$, corrective on-circuit braidings are performed so that the final outcome amounts to applying $U^{(4)}$. More precisely, if the four possibilities for the projective measurements are $\Pi_{21}^{(p_2 p_1)} = \Pi_2^{(p_2)}\Pi_1^{(p_1)}$, with corresponding outcomes  ${(p_2,p_1)\in\{++,+-,-+,--\}}$, these corrections are given by the four possible expansions:
\begin{align}
    & {\rm exp}\left(-i\frac{\pi}{4} \hat\gamma_1^a \hat\gamma_3^a \hat\gamma_1^b \hat\gamma_3^b\right)\ket{\psi}=\\
     \nonumber
    & =  2 \mspace{1.5mu} {\rm exp}\left(-\frac{\pi}{4} \hat\gamma_1^b \hat\gamma_6^b\right) \Pi_{21}^{++} \ket{\psi}\\
    \nonumber
    & =  2i \mspace{1.5mu} {\rm exp}\left(\frac{\pi}{2} \hat\gamma_1^a \hat\gamma_3^a\right) {\rm exp}\left(\frac{\pi}{2} \hat\gamma_1^b \hat\gamma_3^b\right) {\rm exp}\left(-\frac{\pi}{4} \hat\gamma_3^b \hat\gamma_6^b\right) \Pi_{21}^{+-} \ket{\psi}\\
    \nonumber
    & =  2i \mspace{1.5mu} {\rm exp}\left(\frac{\pi}{2} \hat\gamma_1^a \hat\gamma_3^a\right) {\rm exp}\left(\frac{\pi}{2} \hat\gamma_1^b \hat\gamma_3^b\right) {\rm exp}\left(+\frac{\pi}{4} \hat\gamma_3^b \hat\gamma_6^b\right) \Pi_{21}^{-+} \ket{\psi}\\
    \nonumber
    & =  2 \mspace{1.5mu} {\rm exp}\left(+\frac{\pi}{4} \hat\gamma_1^b \hat\gamma_6^b\right) \Pi_{21}^{--} \ket{\psi}
\end{align}
where the condition~\eqref{eq:ancillary_mzm} has been used.

In the end, the ancillary pair of MZMs $\hat \gamma_{5,6}^b$ is unchanged; however, it was essential for the practical implementation of the gate.

\bibliography{Braiding}

\end{document}